# Bilayer Insulator Tunnel Barriers for Graphene-Based Vertical Hot-electron Transistors

S. Vaziri[1], M. Belete[2], E. Dentoni Litta[1], A. D. Smith[1], G. Lupina[3], M. C. Lemme[1,2], and M. Östling[1]

[1]KTH Royal Institute of Technology, School of Information and Communication

Technology, Isafjordsgatan 22, 16440 Kista, Sweden

[2]University of Siegen, Hölderlinstrasse 3, 57076 Siegen, Germany

[3]IHP, Im Technologiepark 25, 15236 Frankfurt (Oder), Germany

Vertical graphene-based device concepts that rely on quantum mechanical tunneling are intensely being discussed in literature for applications in electronics and optoelectronics. In this work, the carrier transport mechanisms in semiconductor-insulator-graphene (SIG) capacitors are investigated with respect to their suitability as the electron emitter in vertical graphene base transistors (GBTs). Several dielectric materials as tunnel barriers are compared, including dielectric double layers. Using bilayer dielectrics, we experimentally demonstrate significant improvements in the electron injection current by promoting Fowler-Nordheim tunneling (FNT) and step tunneling (ST) while suppressing defect mediated carrier transports. High injected tunneling current densities approaching $10^3$ A/cm$^2$ (limited by series resistance), and excellent current-voltage nonlinearity and asymmetry are achieved using a 1 nm-thick high quality dielectric, thulium silicate (TmSiO), as the first insulator layer, and titanium dioxide (TiO$_2$) as a high electron affinity second layer insulator. We also confirm the feasibility and effectiveness of our approach in a full GBT structure which shows dramatic improvement in the collector on-state current density with respect to the previously reported GBTs. The device design and the fabrication scheme have been selected with future CMOS process compatibility in mind. This work proposes a bilayer tunnel barrier approach as a promising candidate to be used in high performance vertical graphene-based tunneling devices.

# Introduction

The rise of the first two-dimensional material, graphene, has led to the investigation of a vast number of potential applications in microelectronics and photonics [1]–[4]. One main focus of the graphene research has been on its integration into conventional devices such as field effect transistors (FETs), where the graphene is used as the channel material [5], [6]. Simultaneously, novel graphene-based architectures and device concepts have been introduced to overcome its intrinsic limitations (such as its lack of a band gap) as well as exploiting its potential for high frequency and possible THz applications [7]. Among these, vertical devices such as graphene base transistors (GBTs) [8]–[10], graphene field effect tunneling transistors [11] and carrier tunneling-based graphene photodetectors [12] are fascinating examples, which have attracted excessive attention due to their promising performance projections for THz applications [13]–[15]. The functionality of these devices is based on quantum mechanical tunneling and hot carrier transport perpendicular to the graphene plane. As a consequence, dielectric tunnel barriers in metal-insulator-graphene (MIG) structures, analogous to well established metal-insulator-metal (MIM) structures, play a crucial role in the operation and performance of vertical graphene-based devices. Note that MIG structures may be replaced by semiconductor- or graphene-insulator-graphene (SIG or GIG) structures.

So far, only a limited number of studies have focused on the integration of graphene and conventional dielectric tunnel barriers [16]–[19]. In addition, while 2D crystal materials like h-BN, can potentially be good tunnel barrier candidates [20], the lack of reproducible high-quality large-scale production methods and their lower integration potential with the CMOS platform compared to established dielectrics puts these into a more embryonic stage. Conventional tunnel barriers like atomic layer deposited dielectrics, in contrast, take advantage of their high process controllability and CMOS compatibility and provide more degrees of freedom in the choice of material for barrier design. Out of these materials, bilayer insulators have shown more promise than single insulators as the tunnel barriers in order to obtain the desired nonlinear and asymmetric current-voltage characteristics in MIM diodes [21]. Note that the term "bilayer" refers to the choice of two dielectric materials, not to a material composed of two stacked monoatomic layers of two-dimensional crystals. In this work, we investigate transport through semiconductor-insulator-insulator-graphene (SIIG) tunnel diodes using atomic layer deposited (ALD) dielectrics including the novel dielectrics $Tm_2O_3$ and TmSiO with respect to their suitability for GBTs. ALD $Tm_2O_3$ is a polycrystalline material with a dielectric constant of about 16 [22]. $Tm_2O_3$ has a bandgap of 6.5 eV and 5.3 eV for MBE on Si [23], and ALD on Ge [24], respectively. The reported conduction and valance band offsets (CBO/VBO) are 2.3 eV/3.1 eV for MBE on Si [23] and 1.7 eV/2.9 eV for ALD on Ge [24]. In addition, a TmSiO layer with a dielectric constant of 12 is formed by rapid thermal annealing (RTA) of ALD $Tm_2O_3$ on Si[25].

The GBT consists of a graphene base electrode, which is separated from emitter and collector electrodes by a tunnel barrier (emitter-base insulator: EBI) and a filtering barrier (base-collector insulator: BCI), respectively (Figure 1a). Figure 1b illustrates the corresponding simplified band diagram of a GBT in the on-state biasing condition. The emitter injects electrons through the EBI tunnel barrier to the graphene base. Thanks to graphene's ultimate thinness, electrons can pass through the graphene to enter the conduction band of the BCI. In order to yield high frequency performance, the emitter current has to meet the following requirements:

1. The current is dominated by injection of hot electrons to the graphene base (tunneling or thermionic emission). When these electrons have energies well above the Fermi level of the graphene base and the collector barrier height, they can overcome the collector barrier and contribute to the collector on-current. This leads to a high current gain of the device.

2. Emitter-base emission of cold electrons should be prevented. Those electrons with energies comparable to the graphene base Fermi level can easily be backscattered from the base-collector barrier and contribute to the undesirable parasitic base current. In this case, the emission of cold electrons can be attributed to defect mediated electron transfer mechanisms and direct tunneling (DT) of the electrons in lower energy levels of Si.

3. A high current density is needed to satisfy the high frequency operation requirement.

4. High nonlinearity is required to obtain a high transconductance.

5. A low threshold voltage is essential for low voltage operation of GBTs.

To satisfy all these requirements, Fowler-Nordheim tunneling (FNT), resonant tunneling (RT), and thermionic emission are the most promising carrier transport mechanisms. In this paper, we focus on dielectric barriers to promote FNT. The difference between FNT and DT lies in the shape of the barrier which electrons encounter. In DT, the electrons tunnel through a trapezoidal barrier, whereas FNT is through a triangular barrier, resulting in higher nonlinearity due to the voltage dependent effective barrier thickness reduction. High tunneling currents should be achieved by using tunnel barriers with very small barrier heights and thicknesses. However, low band gap dielectrics like $Ta_2O_5$ and $TiO_2$ are well-known for their large defect densities preventing dominant tunneling currents or thermionic emission through thin layers of these dielectrics. Bilayers consisting of a high quality dielectric (layer 1) and a low band gap dielectric (layer 2) can efficiently suppress both DT and defect mediated currents and; thus make FNT the dominant transport mechanism (figure 1c). Moreover, utilizing layer 2 dielectrics with very high electron affinity and proper thickness can, in principle, result in step tunneling (ST) [21] (figure 1d), in which the effective barrier thickness is suddenly reduced to the thickness of the layer with the lower electron affinity (layer 1). In this work, several different dielectrics were studied as tunnel barriers for GBTs. Specifically, we utilized atomic layer deposited thulium oxide ($Tm_2O_3$) to form thulium silicate (TmSiO) interlayers which are known to result in well-controlled high quality interfaces to silicon [26], [27]. Finally, we demonstrate that applying a high and low electron affinity insulator stack of TmSiO-$TiO_2$ results in a nonlinear and high-level tunneling current.

## Fabrication

The substrates with patterned emitter and contact areas were prepared on 8-inch n-type antimonide (Sb)-doped (0.01-0.02 Ohm.cm) Si (100) wafers. After cleaning, the wafers were covered with a silicon nitride layer, which served as a hard mask and a stop layer for chemical mechanical polishing (CMP). The active and contact areas were patterned using photolithography and reactive ion etching of $Si_3N_4$ and Si. In the next step, the trenches were filled with high-density plasma undoped silicon glass (HDP USG) and planarized by CMP resulting in a final thickness of the isolation of roughly 650 nm [28]. Subsequently, the contact areas were additionally implanted with As to increase the active dopant concentration to approximately $1x10^{20}$ $cm^{-3}$. After removing $SiO_2$ from the Si pillars, a self-aligned silicidation process was performed to obtain $CoSi_2$ in the exposed regions. The native oxide was removed from the silicon active areas by HF wet etch. Immediately, the samples were loaded into an atomic layer deposition (ALD) reactor to deposit thin film dielectrics. At this step, the experiment was divided into five samples with different dielectric stacks: $Al_2O_3$/$HfO_2$ (2 nm/2 nm), TmSiO/$HfO_2$ (1 nm/3nm), TmSiO/$Tm_2O_3$ (1 nm/2.8 nm), TmSiO/$TiO_2$ (1 nm/5.5 nm), and TmSiO (1nm). The total thicknesses were targeted based on our previous experience in order to achieve high dominant FNT or ST current densities, and to minimize DT and defect mediated carrier transport. In addition to deionized water vapor as the oxidant for all the depositions, the following precursors were employed: $TmCp_3$ for $Tm_2O_3$, $Hf[C_5H_4(CH_3)]_2(OCH_3)CH_3$ for $HfO_2$, $Al(CH_3)_3$ for $Al_2O_3$, and $TiCl_4$ for $TiO_2$. The deposition temperatures were 200 °C, 350 °C, 200 °C and 250 °C respectively. All thicknesses were measured using spectroscopic ellipsometry. For the $Al_2O_3$/$HfO_2$ sample, an ozone treatment step was done on the Si surface prior to the deposition of the $Al_2O_3$ layer, in order to form an interfacial $SiO_2$ layer of approximately 0.5 nm to improve the interface quality. In the samples with the TmSiO layer, in contrast, the layer itself serves as an interfacial layer. This 1-nm silicate layer is formed by ALD deposition of $Tm_2O_3$ and subsequent rapid thermal annealing (RTA) at 500 °C for 1 min. The remaining $Tm_2O_3$ is selectively wet etched in $H_2SO_4$ [27]. The fact that the thickness of the TmSiO layer is dependent only on the annealing temperature allows very precise tuning of the thickness. Another advantage of this technology, in contrast to the $SiO_2$ interfacial layer, is that the post-deposition ozone treatment of the second dielectric layer does not have an effect on the thickness of TmSiO.

After the deposition of the second dielectric layer, graphene grown by chemical vapor deposition (CVD) on copper was transferred onto the substrates using a PDMS-supported transfer method [29][30]. Note that, in our fabrication scheme, the tunnel barrier is formed prior to the graphene transfer. The main reason is that direct ALD of thin high quality dielectric layers on graphene is very challenging due to the non-functional characteristics of the graphene surface [16], [31], [32]. After pattering graphene with $O_2$ plasma, the graphene layer was contacted using metal evaporation and lift-off of titanium (Ti) / platinum (Pt). Finally, in order to improve the interface quality, a forming gas anneal (FGA) was performed at 350 for 30 min. However, in some cases, especially for graphene on $Tm_2O_3$, we experienced degradation of the graphene layer after FGA. Figure 2a shows the top view optical micrograph of the fabricated SIIG structures. The Raman spectrum of the graphene on the substrate (Figure 1a, inset) confirms the performance of the transfer process with no significant defect introduction. To further asses the fabrication steps and confirm the functionality of graphene, the structure was electrically characterized as a field effect transistor using the substrate as the back gate. All the electrical characterization was done in ambient air and at room temperature. Figure 2b shows the transfer characteristics of a GFET with the corresponding 'V' shape ambipolar characteristics which are indicative of graphene. The inset shows the schematic of the fabricated structure labeled as a back gated transistor.

## Results and discussion

The biasing conditions used throughout this work are defined as forward bias when a positive voltage is applied to the graphene metal contact and reverse bias when a negative voltage is applied, as indicated in Fig. 1c and 1d. The first layer of the bilayer tunnel barriers, thulium silicate TmSiO, has a lower electron affinity, high dielectric quality, and a good interface to the silicon emitter. The second layer must be a low band gap dielectric, which is thick enough to block trap-mediated electron transport through the insulator. This configuration can suppress defect-mediated transport and enable FNT or (preferably) ST as the dominant transport mechanism. This should enable high current densities with high nonlinearity. Previously, we reported on a proof of concept GBT with 5nm of $SiO_2$ as the emitter tunnel barrier [9]. Figure 3 compares current-voltage characteristics of the tunnel diodes in the present report and the previously reported $SiO_2$ barrier GBT. For the insulators used in this experiment, the silicon-dielectric conduction band offset decreases starting from $SiO_2$ (CBO: 3.3 eV, electron affinity $\chi$: 0.75 eV) [33], [34] to $Al_2O_3$ (CBO: 2.8 eV, $\chi$: 1.25 eV) [35] to $Tm_2O_3$ (2 eV, $\chi$: 2 eV) [23] to $HfO_2$ (1.5-2 eV, $\chi$: 2.55 eV) [36] and to $TiO_2$ with the lowest band offset (below 1 eV) [37]–[39]. Note that the electron affinity of Si and graphene are 4.05 eV and 4.4 eV, respectively. The threshold voltage, where conduction sets in, of the 5 nm $SiO_2$ sample is approximately 4.5 V. Replacing this with lower barrier heights and thicknesses decreases the threshold voltage and increases the current. This is confirmed by the experimental data in Fig. 3. The dielectric stack of $TmSiO/TiO_2$, which has the thickest barrier of 6.5 nm and the lowest second layer (TiO2) barrier height, results in the highest increase in current density (Figure 3). The samples with $TmSiO/HfO_2$ (not shown) and $TmSiO/Tm_2O_3$ (blue triangles) tunnel barriers exhibit very similar characteristics.

Several potential transport mechanisms through the double insulator barriers are considered, namely Frenkel-Poole Emission (FPE), DT, FNT, and ST. In the FPE model, current has a voltage and temperature dependency as described by equation 1 [40]

$$J_{FPE} \propto V \exp\left[\frac{q}{KT}(2A\sqrt{V} - \Phi_B)\right] \qquad (1)$$

where V is the voltage drop across the insulator, $\Phi_B$ is the barrier height between the trap energy level and the edge of the dielectric conduction band, q is the elementary charge, T is the temperature, K is the Boltzmann constant, and A is a constant. Equation 1 leads to a linear behavior when the data is plotted as J/V vs. $V^{1/2}$. In forward bias, which is the typical operation range for GBTs, most of the samples show poor linear fits to the FPE model except for a very limited and low voltage ranges in some of the samples (not shown). FPE can therefore be excluded as the dominant conduction mechanism. Only $Al_2O_3/HfO_2$ sample shows a good linear fit to the FPE model for voltages below 4 V. The exclusion of FPE can be further confirmed by temperature dependent I-V measurements (Figure 4a and 4b). If the transport is dominated by tunneling, the temperature I-V characteristics I(T)-V should not have significant temperature dependency. Trap-mediated transport or FPE, in contrast, exponentially depends on the temperature. While in forward bias no significant temperature dependency can be seen, some temperature dependency is observed in the devices with $Al_2O_3/HfO_2$ (in forward bias) and $TmSiO/TiO_2$ (in reverse bias). Moreover, for TmSiO/HfO$_2$ and TmSiO/TiO$_2$ in forward bias, as the temperature dependency decreases in the high field range, we thus expect tunneling to become the dominant transport mechanism at higher voltages.

Tunneling as the dominant transport mechanism can be confirmed by fitting the voltage dependence of the measured current to the FN model [40]:

$$J_{FNT} \propto V^2 \exp\left[\frac{-b}{V}\right] \qquad (2)$$

in which b is constant. Note that the distinction between FNT and DT can typically be made only by considering the thickness of the tunnel barrier and the applied voltage range. The data, plotted as J/V$^2$ vs. V$^{-1}$ in forward bias, shows excellent linearity for SiO$_2$, Al$_2$O$_3$/HfO$_2$, TmSiO/TiO$_2$, and TmSiO samples (Figure 5), with a slightly smaller R$^2$ value for TmSiO/HfO$_2$ and TmSiO/Tm$_2$O$_3$. Here, increasing HfO$_2$ and Tm$_2$O$_3$ thicknesses are expected to result in better fits to the FN model, but an increased thickness would also exponentially reduce the tunneling current, contrary to the desired outcome.

Based on the thickness of the pure TmSiO layer of 1 nm, we are confident that DT is the dominant tunneling mechanism. However, for TmSiO/TiO$_2$, 6.5 nm is too thick for direct tunneling. This can be confirmed by comparing the asymmetry in the I-V characteristics of these samples, defined as the ratio of currents in forward and reverse biasing conditions ($|I_+/I_-|$). Asymmetry can originate from different work functions of the metals, especially if the transport is based on tunneling or Schottky emission. Simultaneously, bilayer tunnel barriers with different electron affinity, dielectric constant, and thicknesses of the two dielectrics introduce asymmetry due to the different transport mechanisms or barriers seen by the carriers travelling in opposite directions. Figure 6a compares the asymmetry of the samples with TmSiO, TmSiO/Tm$_2$O$_3$, TmSiO/TiO$_2$, and Al$_2$O$_3$/HfO$_2$. Very low asymmetry observed in the TmSiO sample is in line with the direct tunneling mechanism deduced from the fit in Figure 5. The different polarity of the asymmetry in TmSiO/Tm$_2$O$_3$ can be attributed to larger reverse bias current due to defect enhanced direct tunneling in the reverse bias condition. In this case, electrons travel from graphene through the Tm$_2$O$_3$ layer via FPE. At the interface of TmSiO/Tm$_2$O$_3$, the electrons directly tunnel through the TmSiO. This defect enhanced direct tunneling has been also shown for the case of Al$_2$O$_3$/Ta$_2$O$_5$ in [41]. Finally, due to the large asymmetry in the barrier shape, TmSiO/TiO$_2$ and Al$_2$O$_3$/HfO$_2$ show highest asymmetry (>1000) which rules out the possibility of dominant direct tunneling and defect mediated transport in both forward and reverse biasing conditions.

Following the discussion so far, two possibilities remain for the TmSiO/TiO$_2$ tunnel barrier in the forward biasing condition: FNT and ST. In bilayer tunnel barriers, step tunneling occurs as a result of a sudden reduction of the barrier thickness when the second barrier "disappears" due to a large difference in the electron affinity of the two dielectrics (Figure 1d). Considering the low band offset of TiO$_2$ with respect to silicon, we propose that ST contributes significantly to the total current in these devices. This can be elucidated by investigating the nonlinearity of the I-V characteristics, (dI/dV)/(I/V). I-V characteristics with dominant FNT or ST mechanisms should show higher nonlinearity in comparison with other carrier transport mechanisms due to the voltage dependent barrier thickness reduction in these mechanisms. Figure 5b compares the nonlinearity for three capacitors at lower voltages (TmSiO, TmSiO/HfO$_2$, and TmSiO/TiO$_2$), with the TmSiO/TiO$_2$ barrier showing the highest nonlinearity. This high nonlinearity at low voltages for a 6.5 nm-thick barrier is in line with the proposed step tunneling carrier transport mechanism through TmSiO/TiO$_2$ barriers.

Further evidence for the absence of trap mediated transport through the TmSiO/TiO$_2$ barrier is shown in Figure 7a: A double voltage sweep reveals almost no hysteresis in the currents. This high quality dielectric barrier can result in very high injected current densities in the order of 10$^3$ A/cm$^2$ (Figure 7b) without hard dielectric breakdown. In addition, the current density in the TmSiO/TiO$_2$ scales with device area (Figure 7b inset). Note that the high current densities in these samples are achieved despite being limited by the series resistance. This can be inferred from the change in the slope of the I-V characteristics shown in figure 7b as well as the deviation in the I-V characteristics for devices with different area aspect ratios in figure 7c. This figure compares the I-V characteristics of two devices with the same area and different aspect ratio between the width and the length of the Si active area (inset). Above 3 volts, the current densities start to slightly deviate from each other, which may be attributed to the difference in the series resistance and potentially current crowding in the active areas with different aspect ratios.

In order to confirm the effectiveness of the TmSiO/TiO$_2$ emitter barriers, a full GBT structure was fabricated. Following the materials proposed in [42], 60 nm of Si was deposited as the collector barrier on top of the graphene for the samples with the TmSiO/TiO$_2$ tunnel barriers. After the formation of the collector electrode using a lift-off process, the Si BCI was patterned applying photolithography and a wet etch process. Figure 8 shows the transfer characteristics of this GBT at a base-collector voltage of $V_{BC} = 0$. Note that higher collector bias leads to substantial leakage currents due to the non-optimized BCI deposition process. We are nevertheless including this preliminary data with a focus on the emitter barrier, while the BCI optimization is beyond the scope of this article. Setting $V_{BC} = 0V$ avoids direct leakage between the base and the collector and allows investigating the hot-electron transport. Even at zero base-collector bias, this GBT with a step tunnel barrier shows orders of magnitude higher on-state current density and an improved current transfer ratio $\alpha$ ($I_C/I_E$) of more than 20% in comparison to the previously reported GBT with an $SiO_2$ tunnel barrier and an $\alpha$ of approximately 6%[9].

## Conclusion

In conclusion, we have investigated the feasibility of bilayer tunnel dielectrics as injectors for hot electrons in graphene base transistors. Table 1 summarizes the tunneling barrier characteristics investigated in this work. We demonstrated the application of thulium silicate as a high quality interfacial dielectric layer which also serves as the first tunnel barrier in the bilayer tunnel stack. High tunneling current densities of $10^3$ A/cm$^2$ with excellent nonlinearity were achieved using this high quality dielectric layer in conjunction with $TiO_2$ as the second layer. This high electron affinity dielectric suppresses defect mediated carrier transport and the injection of the cold electrons via direct tunneling, and instead promotes Fowler-Nordheim and step tunneling. The results show dramatic improvement in the injection current with respect to the reference $SiO_2$ tunnel barrier. In addition, applying the bilayer tunnel stack of $TmSiO/TiO_2$ in a GBT with a Si BCI layer resulted in orders of magnitude larger collector current density with respect to the original data for the GBT with the $SiO_2$ EBI. Moreover, the proposed materials, device design, and fabrication scheme enable the repeatable and scalable exploration of the performance limits and further optimization of the bilayer tunnel barriers for high performance graphene base hot electron transistors.

**Table 1 Dielectric materials used as the tunnel barriers and the corresponding carrier transport characteristics.**

| Tunneling Barrier | Thickness (nm) | I-V characteristics | Dominant transport mechanism |
|---|---|---|---|
| TmSiO | 1 | • Minor T-dependency<br>• Excellent fit to FN model: $R^2=0.999$<br>• Poor fit to FPE in the forward bias<br>• Low asymmetry: <10<br>• Low nonlinearity | Direct Tunneling |
| TmSiO/TiO$_2$ | 1/5.5 | • Minor T-dependency (especially at high field range) in the forward bias, high temperature dependency in the reverse bias<br>• Excellent fit to FN model: $R^2=0.999$<br>• Poor fit to FPE in the forward bias<br>• High asymmetry: $>10^3$<br>• High nonlinearity | Fowler-Nordheim Tunneling/Step Tunneling in the forward bias<br>defect mediated transport in the reverse bias |
| TmSiO/Tm$_2$O$_3$ | 1/2.8 | • Minor T-dependency (especially at high field range)<br>• good fit to FN model: $R^2=0.996$<br>• Poor fit to FPE in the forward bias<br>• Moderate asymmetry: >50<br>• Moderate nonlinearity | Fowler- Nordheim Tunneling in the forward bias |
| TmSiO/HfO$_2$ | 1/3 | • Minor T-dependency (especially at high field range)<br>• good fit to FN model: $R^2=0.992$<br>• Poor fit to FPE in the forward bias<br>• Moderate asymmetry: >50<br>• Moderate nonlinearity | Fowler- Nordheim Tunneling at higher electric fields |
| Al$_2$O$_3$/HfO$_2$ | 2/2<br>(4.5 nm including the SiO$_2$ interfacial layer) | • Minor T-dependency in the reverse bias, higher T-dependency in the forward bias<br>• good fit to FN model: $R^2=0.999$<br>• good fit to FPE in the forward bias for v<4 V: $R^2=0.999$<br>• High asymmetry: $>10^3$<br>• High nonlinearity | Frenkel-Poole emission in the forward bias condition <4 V<br>Fowler-Nordheim at higher fields |
| SiO$_2$ | 5 | • Minor T-dependency<br>• Excellent fit to FN model: $R^2=0.998$ | Fowler- Nordheim Tunneling |

**Acknowledgements**

The authors thank F. Driussi, S. Venica, P. Palestri, and L. Selmi (Univ. Udine) for fruitful discussions. Support from the European Commission through a STREP project (GRADE, No. 317839), the Swedish Research Council and an ERC Starting Grant (InteGraDe, No. 307311) as well as the German Research Foundation (DFG, LE 2440/1-1) is gratefully acknowledged. KTH Excellence Ph.D. Scholarship is also acknowledged.

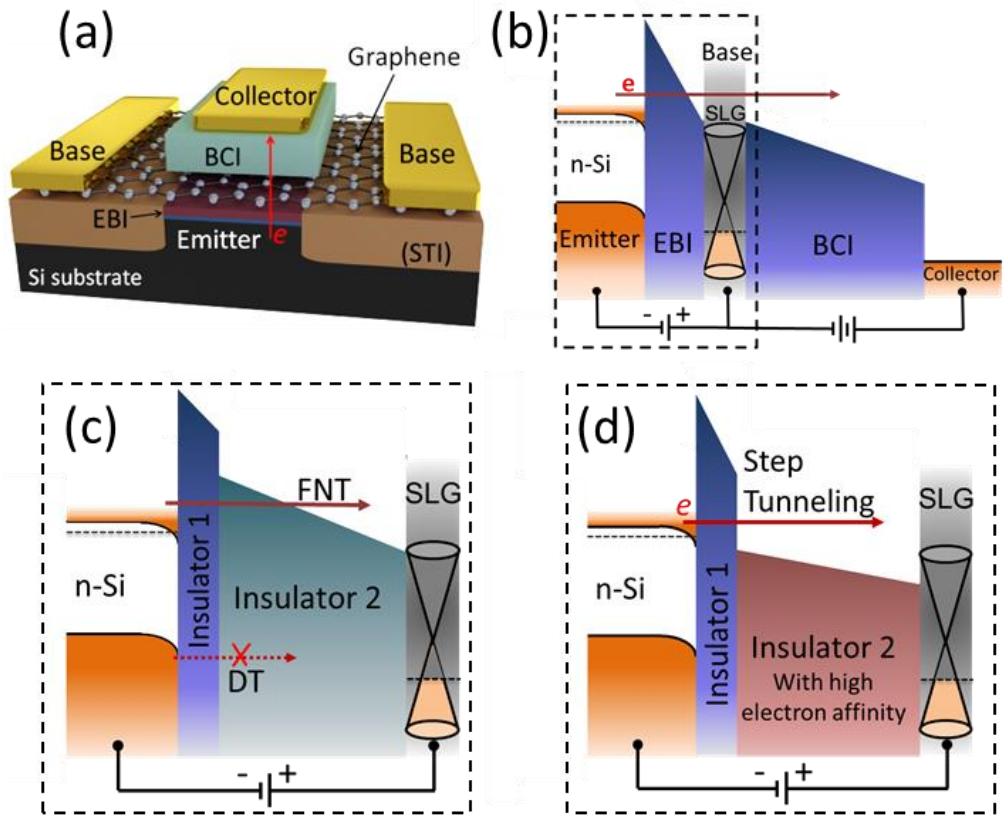

**Figure 1:** (a) Schematic isometric view of the GBT. The red arrow indicates electrons transport direction. (b) Simplified band diagram of the GBT in the on-state. (c) The injection diode (dashed rectangular in b) with bilayer insulator stack showing Fowler-Nordheim Tunneling (FNT). (d) The same injection diode as c but with a higher electron affinity insulator 2, showing step tunneling (ST).

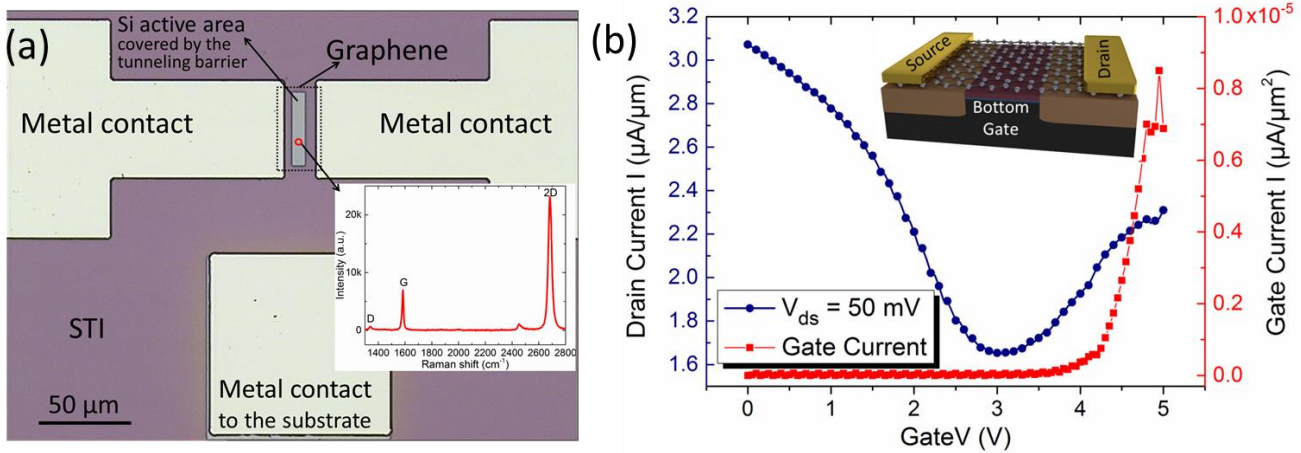

**Figure 2:** (a) Top-view micrograph of a fabricated device with the graphene Raman fingerprint shown as an inset. The Raman spectrum confirms the presence of the single layer graphene with very small in-plane crystal defect

related peak (D peak). (b) Transfer characteristics (blue circles) and gate leakage current (red squares) of a GFET with 5 nm $Al_2O_3$/$HfO_2$ gate dielectric stack. The inset shows the schematic of the back gated GFET.

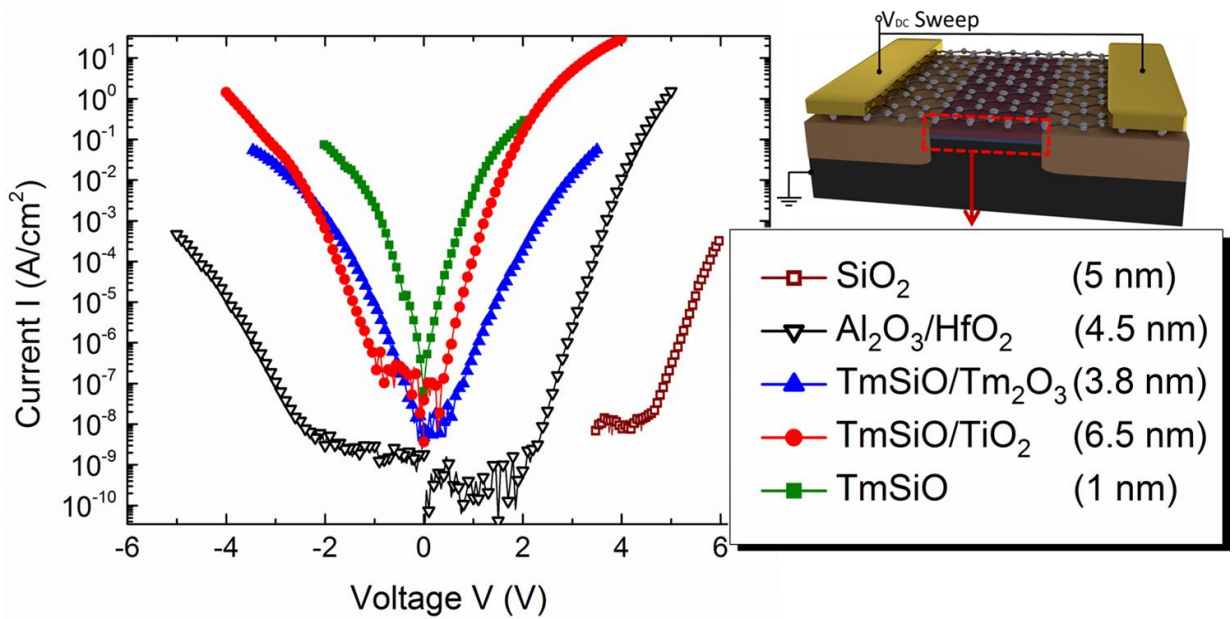

**Figure 3:** I-V characteristics of the SIIG tunnel diodes with different tunnel barrier stacks. The schematic illustrates the biasing condition and highlights the area of the tunneling insulators in the devices. Devices with bilayer insulators, which combine the high quality interface layer of TmSiO with a second insulator with higher electron affinity (this work) with respect to $SiO_2$ (Ref. 9), show superior I-V characteristics. $TmSiO/TiO_2$ tunneling stacks show particularly promising characteristics: low threshold voltage, high current, and high nonlinearity.

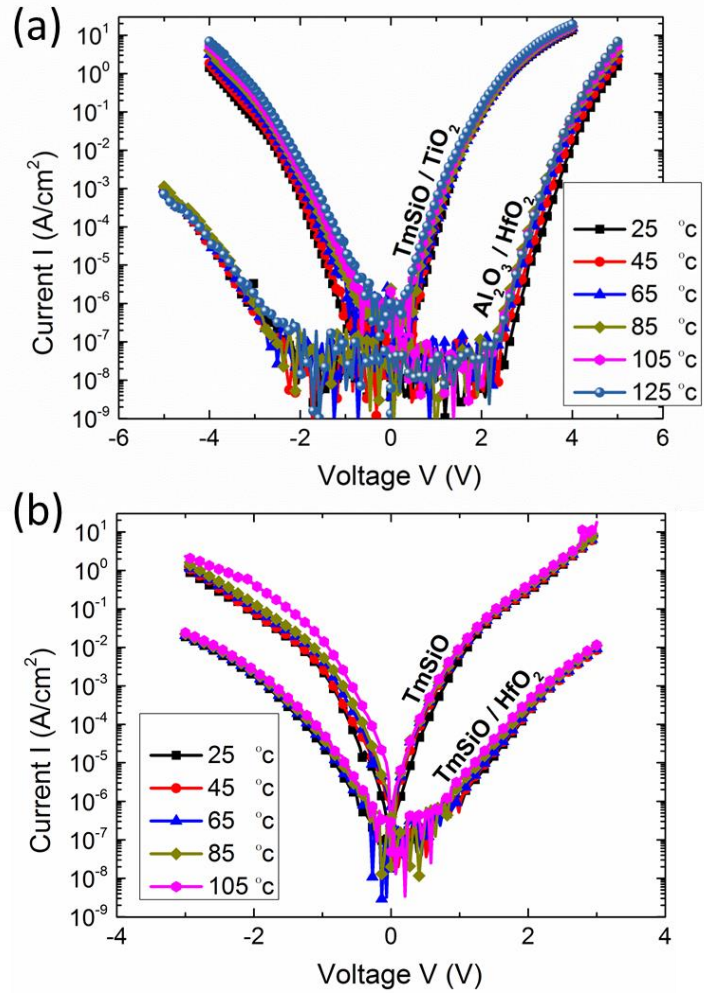

**Figure 4:** Temperature dependent I-V characteristics of the tunnel diodes with (a) TmSiO/TiO$_2$ and Al$_2$O$_3$/HfO$_2$ and (b) TmSiO and TmSiO/HfO$_2$ tunnel barriers. Some temperature dependency can be observed for TmSiO/TiO$_2$ in the reverse bias and Al$_2$O$_3$/HfO$_2$ in the forward bias conditions. In the forward bias, TmSiO/TiO$_2$ and TmSiO/HfO$_2$ only exhibit a very small temperature dependency limited to very low voltages. This temperature dependency diminished in higher voltages by the domination of tunneling mechanism.

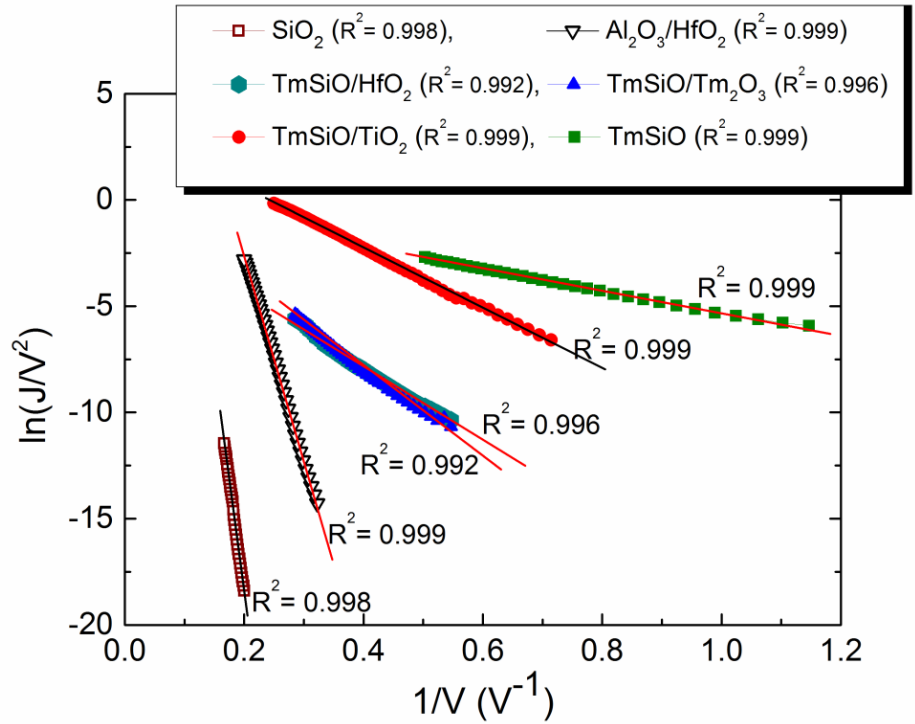

**Figure 5:** Fowler-Nordheim plots in the forward biasing condition shows excellent linear behavior in the tunnel barriers, a strong evidence of F-N tunneling.

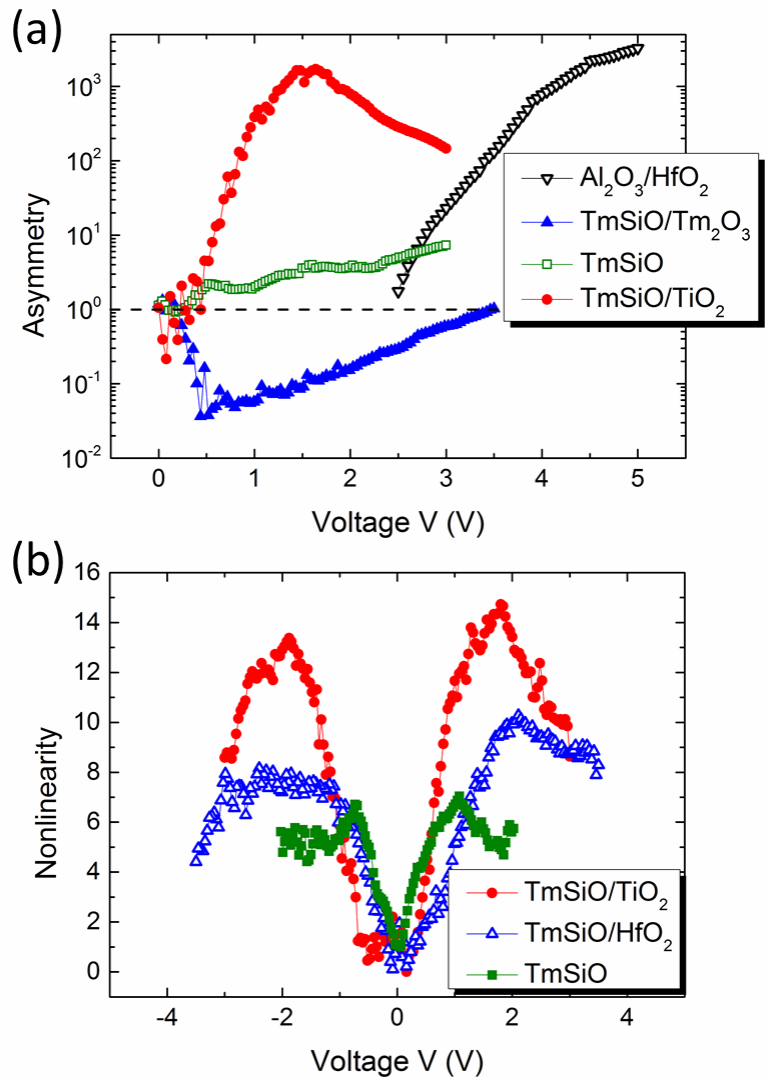

**Figure 6:** **(a) Asymmetry plot shows limited values for 1 nm TmSiO tunnel barrier which is an indication of direct tunneling. In the case of TmSiO/TiO$_2$ asymmetries more than 1000 are observed due to the asymmetry of the bilayer tunnel barrier. (b) I-V nonlinearity in TmSiO, TmSiO/HfO$_2$, and TmSiO/TiO$_2$ tunnel barriers. TmSiO/TiO$_2$ shows the highest nonlinearity which may be attributed to the voltage dependent barrier thickness reduction.**

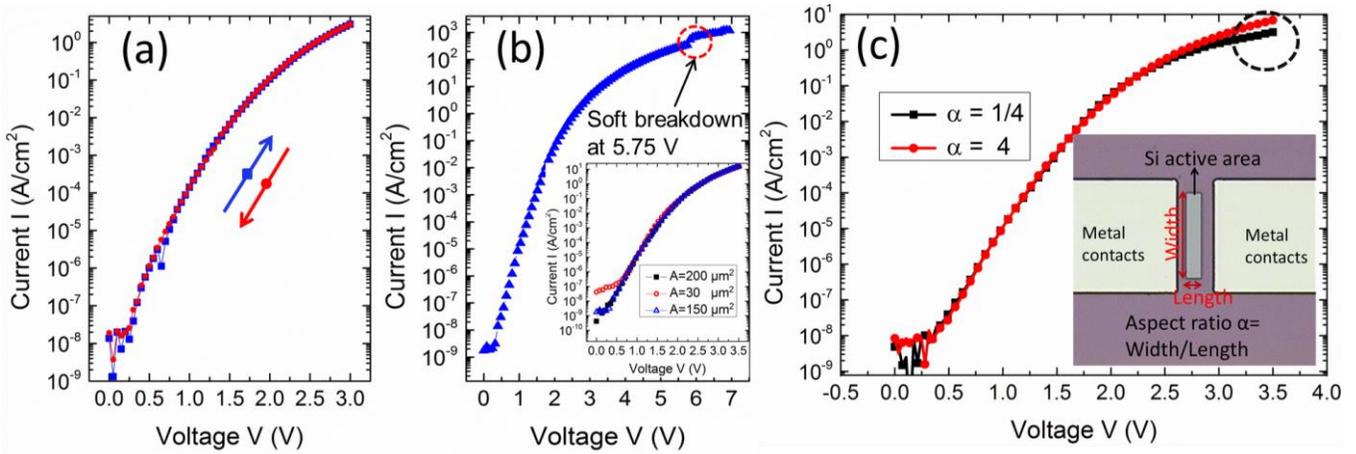

**Figure 7:** (a) Forward / reverse current-voltage measurements for a tunneling diode with a TmSiO/TiO$_2$ dielectric stack. No significant hysteresis is observed. (b) I-V characteristics of a TmSiO/TiO$_2$ tunnel diode in forward bias in a large voltage range. High current density of $10^3$ A/cm$^2$ was achieved with soft breakdown at 5.75 V. The inset shows that the current scales with the area. (c) Comparison of the I-V characteristics with the same dielectric stack as (a) and two different aspect ratios and identical areas. Lower current densities for the device with lower aspect ratio is emphasized with a dashed circle.

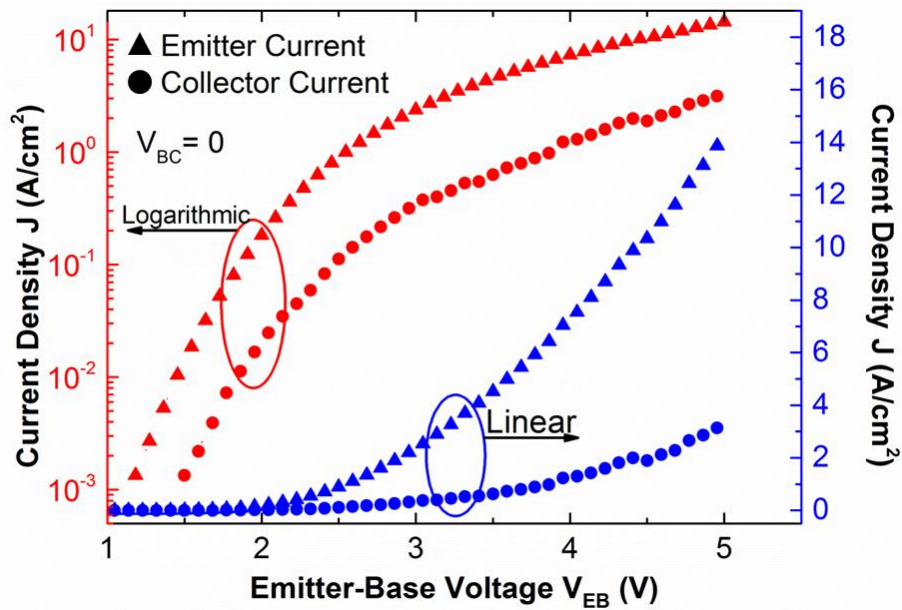

**Figure 8:** Transfer characteristics of a GBT with the TmSiO/TiO$_2$ emitter tunnel barrier and 60 nm deposited Si as the BCI. The left and right-hand axes (red and blue) show the current in logarithmic and linear scales, respectively. While the triangles show the emitter current the circles display the collector current. The collector current has dramatically improved in comparison with previously reported GBTs with the on-state current densities in the order of 10 μA/cm$^2$ [9].